\documentclass[11pt,a4paper]{article}
\usepackage{graphicx}
\usepackage{jcappub}

\def\sun{\hbox{$\odot$}}

\def\la{\mathrel{\mathchoice {\vcenter{\offinterlineskip\halign{\hfil
$\displaystyle##$\hfil\cr<\cr\sim\cr}}}
{\vcenter{\offinterlineskip\halign{\hfil$\textstyle##$\hfil\cr
<\cr\sim\cr}}}
{\vcenter{\offinterlineskip\halign{\hfil$\scriptstyle##$\hfil\cr
<\cr\sim\cr}}}
{\vcenter{\offinterlineskip\halign{\hfil$\scriptscriptstyle##$\hfil\cr
<\cr\sim\cr}}}}}
\def\ga{\mathrel{\mathchoice {\vcenter{\offinterlineskip\halign{\hfil
$\displaystyle##$\hfil\cr>\cr\sim\cr}}}
{\vcenter{\offinterlineskip\halign{\hfil$\textstyle##$\hfil\cr
>\cr\sim\cr}}}
{\vcenter{\offinterlineskip\halign{\hfil$\scriptstyle##$\hfil\cr
>\cr\sim\cr}}}
{\vcenter{\offinterlineskip\halign{\hfil$\scriptscriptstyle##$\hfil\cr
>\cr\sim\cr}}}}}

\title{An independent  constraint on the secular rate  of variation of
  the gravitational constant from pulsating white dwarfs}
  
\author[a,b]{Alejandro H. C\'orsico,}
\author[a,b]{Leandro G. Althaus,}
\author[c,d]{Enrique Garc\'{\i}a--Berro,}
\author[e]{and Alejandra D. Romero}

\affiliation[a]{Facultad de Ciencias Astron\'omicas y Geof\'{\i}sicas,  
                Universidad  Nacional de La Plata,\\
                Paseo del  Bosque s/n,  
               (1900) La Plata, 
                Argentina}
\affiliation[b]{Instituto de Astrof\'{\i}sica La Plata, 
                CONICET-UNLP, Argentina}
\affiliation[c]{Departament de F\'\i sica Aplicada, 
                Universitat Polit\`ecnica de Catalunya,\\ 
                c/Esteve Terrades, 5,  
                08860 Castelldefels,  
                Spain}
\affiliation[d]{Institute for Space Studies of Catalonia,\\
                c/Gran Capit\`a 2--4, Edif. Nexus 104,   
                08034  Barcelona, 
	        Spain}
\affiliation[e]{Departamento de Astronomia,
                Universidade Federal do Rio Grande do Sul,\\
                Av. Bento Goncalves 9500, Porto Alegre 91501-970, RS, 
                Brazil}

\emailAdd{acorsico@fcaglp.unlp.edu.ar}

\abstract{A secular  variation of the gravitational  constant modifies
  the structure and  evolutionary time scales of  white dwarfs.  Using
  an  state-of-the-art stellar  evolutionary  code  and an  up-to-date
  pulsational code we  compute the effects of a  secularly varying $G$
  on the  pulsational properties  of variable white  dwarfs. Comparing
  the the theoretical results obtained taking into account the effects
  of a  running $G$ with  the observed  periods and measured  rates of
  change of  the periods of  two well studied pulsating  white dwarfs,
  G117--B15A and R548,  we place constraints on the  rate of variation
  of  Newton's  constant.  We  derive  an  upper bound  $\dot  G/G\sim
  -1.8\times  10^{-10}$~yr$^{-1}$  using   the  variable  white  dwarf
  G117--B15A, and  $\dot G/G\sim -1.3\times  10^{-10}$~yr$^{-1}$ using
  R548.  Although  these upper  limits are currently  less restrictive
  than those obtained using other  techniques, they can be improved in
  a future measuring the rate of change of the period of massive white
  dwarfs.}

\notoc

  
\begin{document}

\maketitle  

%______________________________________________________________________ 
   
\section{Introduction}  
\label{intro}  

The  most  commonly accepted  theory  of  gravitation, namely  General
Relativity, relies on the  assumption that the gravitational constant,
$G$,  is  indeed an  universal  and  genuine  constant.  However,  the
assumption that  $G$ does  not vary  with time or  location is  just a
hypothesis, though quite a reasonable an important one, which needs to
be    observationally    corroborated.     Actually,    some    modern
grand-unification theories predict  that the gravitational constant is
a  slowly  varying  function   of  low-mass  dynamical  scalar  fields
\cite{LAea,U,mio}. Thus, should these theories prove to be correct, it
is  foreseen that the  gravitational constant  would vary  slowly over
long timescales.

In recent years,  the issue of the variation  of fundamental constants
has experienced a renewed  interest, and several observational studies
have  scrutinized  their hypothetical  variations.   The  target of  a
sizable  fraction of  these  studies \cite{U,mio}  has  been to  place
constraints on the variation of the fine structure constant, $\alpha$,
whereas other  studies focused on the  variation of proton-to-electron
mass ratio,  $\mu$.  This stems from  the fact that  the properties of
electromagnetic radiation are very  sensitive to the precise values of
these constants. In sharp contrast with the multitude of studies about
whether  (or not)  there  is  evidence for  a  varying fine  structure
constant or a varying proton-to-electron  mass ratio, only a few works
have  been   devoted  to  study   a  hypothetical  variation   of  the
gravitational constant.  Most  likely, one of the reasons  for this is
the  intrinsic experimental difficulties  involved in  determining the
value of this constant \cite{MTN} on  Earth.  As a matter of fact, the
accuracy of this measurement is orders of magnitude lower than that of
other fundamental  constants, like $\alpha$  or $\mu$, making  $G$ the
fundamental constant with the poorest determination. Being this reason
important,  there  are  as  well  other problabe  reasons,  being  the
weakness of  the gravitational interaction  one of them. Hence,  it is
not surprising the large variety of methods aimed to place limits on a
possible  variation of  $G$.  Among  them the  most  restrictive upper
limits  are  those  obtained  using  Lunar  Laser  Ranging,  Big  Bang
nucleosynthesis  and  the  luminosity  function  of  white  dwarfs  in
clusters.

Lunar  Laser  Ranging has  provided  an upper  bound  to  the rate  of
variation of the gravitational constant $\dot{G}/G = (0.2\pm0.7)\times
10^{-12}$~yr$^{-1}$   \cite{H10},  but   it  is   purely   local.   At
cosmological distances the tightest bound  to the rate of variation of
$G$ is that obtained from  Big Bang nucleosynthesis.  This bound turns
out   to  be   of   the   same  order   of   magnitude  $-0.3   \times
10^{-12}$~yr$^{-1}  \la  \dot{G}/G  \la 0.4\times  10^{-12}$~yr$^{-1}$
\cite{CO4,B05}, but unfortunately is model-dependent.  At intermediate
ages  the  Hubble diagram  of  Type Ia  supernovae  has  been used  to
constrain  the rate  of  variation of  $G$,  yielding somewhat  weaker
constraints,  $\dot  G/G\la   1\times  10^{-11}$~yr$^{-1}$  at  $z\sim
0.5$~\cite{SNIa,IJMPD}.

White dwarfs can be used to set independent bounds on any hypothetical
variation of $G$.  Perhaps, the most important reason for this is that
white    dwarfs    have     very    long    evolutionary    timescales
\cite{review}. Hence, even small rates  of change of $G$ leave sizable
imprints in their evolutionary properties. Nevertheless, there as well
other  reasons, which  are  at least  as  important as  this one.   In
particular,  white dwarfs represent  the final  state of  evolution of
stars  with masses smaller  than $\sim  10\,M_{\odot}$.  Consequently,
most of the stars will end up their lives as a white dwarf.  Moreover,
the mechanical structure of the  inner cores of white dwarfs is almost
entirely supported  by degenerate  electrons. Thus, this  structure is
very sensitive  to the precise value  of $G$, and any  small change in
its value can become apparent.  Additionally, white dwarf stars do not
have  nuclear  energy sources.   Thus,  their  evolution  can be  well
understood  in  terms   of  a  slow  cooling  process   in  which  the
gravothermal energy of the core  is radiated through a thin, partially
degenerate, insulating convective or radiative envelope.  This cooling
process is now  very well understood for the  luminosities of interest
\cite{nature}.  Additionally,  it has been  recently shown \cite{gnew}
that  the  specific  rate at  which  white  dwarfs  cool is  not  only
sensitive  to  its  secular  rate of  variation,  confirming  previous
theoretical evidence  \cite{gold}, but also  to the specific  value of
$G$ at which the white dwarf was born \cite{JCAP}.

In  this paper we  use white  dwarfs to  place an  upper limit  to the
secular rate  of change of the  gravitational constant. To  do this we
use a direct  measure of the evolutionary timescales  of two pulsating
white dwarfs, G117$-$B15A  and R548. These two white  dwarfs belong to
the class  of ZZ  Ceti stars  --- also known  as DAV  pulsators. These
stars are  pulsating white dwarfs with  hydrogen-rich envelopes.  They
have been monitored  for very long periods of  time (decades) and have
good determinations of the periods and rates of period change of their
main pulsational  modes.  In particular, G117$-$B15A  is the benchmark
member   of  the   ZZ  Ceti   class,  its   pulsation   periods  being
$\Pi=215.20$~s, 270.46~s and 304.05~s \cite{kepler82}. The most recent
determination of  the rate of  change of the  period at 215~s  --- its
main  pulstion  period  --- of  this  star  is  $\dot \Pi  =  (4.19\pm
0.73)\times 10^{-15}$~s~s$^{-1}$  \cite{kepler12}.  R548 ---  that is,
ZZ  Ceti itself  --- is  another DAV  star that  has  been intensively
studied for the  last few decades.  Both G117$-$B15A  and R548 display
similar pulsational properties, share periods in the intervals ranging
from  213 to  215~s and  from 272  to 274~s,  and are  expected  to be
similar in  structure.  Since the  discovery of R548, there  have been
multiple attempts to measure the drift rate of its pulsation period at
$\sim 213.13$~s,  but only  very recently the  rate of change  of this
period  with time  has  been measured  for  the very  first time.   In
particular, using 41 years of time-series photometry from 1970 to 2011
a value of $\dot  \Pi=(3.3\pm 1.1)\times 10^{-15}$~s~s$^{-1}$ has been
obtained~\cite{Mukadam12}.

Our paper is organized as follows.  In Sect.  \ref{astero} we describe
the asteroseismological models of the  two pulsating white dwarfs used
in this study.  We discuss extensively the effects of a running $G$ in
the periods and  periods derivatives in Sect.   \ref{method}, while in
Sect.~\ref{G} we use these results  to place constraints on a possible
variation of the gravitational constant.  Finally, in Sect. \ref{conc}
we  summarize our  major  findings, we  discuss  its significance  and
present our concluding remarks.

%______________________________________________________________________ 

\section{Asteroseismological models for G117$-$B15A and R548}
\label{astero}

Here,    we   briefly    describe   the    characteristics   of    the
asteroseismological models  for G117$-$B15A  and R548 employed  in our
analysis and  obtained in a previous  work \cite{2012MNRAS.420.1462R}.
We refer  the interested reader to  that paper for further  details of
the  method.   Briefly,  a thorough  asteroseismological  analysis  of
G117$-$B15A and R548 (and 42 other  bright DAV stars) was performed in
\cite{2012MNRAS.420.1462R} using a grid of fully evolutionary DA white
dwarf models  characterized by  consistent chemical profiles  for both
the  core and  the  envelope, and  covering a  wide  range of  stellar
masses,   thicknesses  of   the   hydrogen   envelope  and   effective
temperatures.   These  models were  generated  with  the {\tt  LPCODE}
evolutionary   code   \cite{2005A&A...435..631A}.   The   evolutionary
calculations   were   carried   out   from  the   ZAMS   through   the
thermally-pulsing and mass-loss phases on  the AGB, and finally to the
domain  of planetary  nebulae  and white  dwarfs.   These  models
  represent  a significant  improvement over  previous ones,  that use
  simplified  chemical  profiles  in  the  envelope  and/or  the  core
  \cite{2010ApJ...717..897A,  2012MNRAS.420.1462R}.    The  effective
temperature, the stellar mass and the mass of the H envelope of our DA
white dwarf models vary within the ranges $14\,000 \ga T_{\rm eff} \ga
9\,000$ K, $0.525  \la M_* \la 0.877 M_{\sun}$,  $-9.4 \la \log(M_{\rm
  H}/M_*) \la -3.6$, where the value of the upper limit of $M_{\rm H}$
is dependent on  $M_*$ and fixed by prior  evolution.  For simplicity,
the  mass  of  He  was  kept  fixed at  the  value  predicted  by  the
evolutionary computations for each sequence.

\begin{table}
\centering
\caption{Characteristics  of  G117$-$B15A  derived from  spectroscopic
  analysis  and results  of the  best asteroseismological  model.  The
  ranges of  values in the second  column have been  derived by taking
  into       account        several       spectroscopic       analysis
  \cite{1995ApJ...438..908R, 2000BaltA...9..119K, 2001ASPC..226..299K,
  1995ApJ...449..258B, 2004ApJ...600..404B}.  The quoted uncertainties
  in  the asteroseismological  model are  the internal  errors  of our
  period-fitting procedure.}
\begin{tabular}{lcc}
\hline
\hline
 Quantity                        & Spectroscopy      & Asteroseismology                              \\
\hline
$T_{\rm eff}$ [K]                & $11\,430-12\,500$ & $11\,985 \pm 200$                             \\
$M_*/M_{\odot}$                  & $0.530-0.622$     & $0.593 \pm 0.007$                             \\
$\log g$                         & $7.72-8.03$       & $8.00\pm 0.09$                                \\
$\log(R_*/R_{\odot})$            &    ---            & $-1.882\pm 0.029$                             \\   
$\log(L_*/L_{\odot})$            &    ---            & $-2.497 \pm 0.030$                            \\
$M_{\rm He}/M_*$                 &    ---            & $2.39 \times 10^{-2}$                         \\
$M_{\rm H}/M_*$                  &    ---            & $(1.25\pm 0.7) \times 10^{-6}$                \\
$X_{\rm C},X_{\rm O}$ (center)   &    ---            & $0.28^{+0.22}_{-0.09} , 0.70^{+0.09}_{-0.22}$ \\
\hline
\hline
\end{tabular}\\
\label{tab:G117-d}
\end{table}

\begin{table}  
\centering  
\caption{Same  as Table  \ref{tab:G117-d},  but for  R548. The  second
  column contains  the spectroscopically determined  values of $T_{\rm
  eff}$ and $\log g$  \cite{2004ApJ...600..404B}, and the stellar mass
  \cite{2012MNRAS.420.1462R}.}
\begin{tabular}{lcc}  
\hline  
\hline  
 Quantity                        & Spectroscopy      & Asteroseismology                              \\    
\hline  
$T_{\rm eff}$ [K]                & $11\,990 \pm 200$ & $11\,627 \pm 390$                             \\  
$M_*/M_{\sun}$                   & $0.590 \pm 0.026$ & $0.609 \pm 0.012$                             \\  
$\log g$                         & $7.97 \pm 0.05$   & $8.03\pm 0.05$                                \\  
$\log(R_*/R_{\sun})$             &    ---            & $-1.904\pm 0.015$                             \\  
$\log(L_*/L_{\sun})$             &    ---            & $-2.594 \pm 0.025$                            \\  
$M_{\rm He}/M_*$                 &    ---            & $2.45 \times 10^{-2}$                         \\  
$M_{\rm H}/M_*$                  &    ---            & $(1.10\pm 0.38) \times 10^{-6}$               \\  
$X_{\rm C},X_{\rm O}$ (center)   &    ---            & $0.26^{+0.22}_{-0.09} , 0.72^{+0.09}_{-0.22}$ \\  
\hline  
\hline  
\end{tabular}  
\label{tab:R548-d}  
\end{table}  

In  order to  find an  asteroseismological model  for  G117$-$B15A and
R548, we searched for the models that minimize a quality function that
measures the distance between theoretical ($\Pi^{\rm t}$) and observed
($\Pi^{\rm  o}$) periods \cite{2012MNRAS.420.1462R}.   The theoretical
periods were  assessed by means of an  up-to-date adiabatic, nonradial
pulsation  code \cite{2006A&A...454..863C}  that solves  the equations
for   linear   non-radial   stellar   pulsations  in   the   adiabatic
approximation \cite{1989nos..book.....U}.   Pulsations in white dwarfs
are associated to nonradial $g$(gravity)-modes which are a subclass of
spheroidal modes  whose main restoring  force is gravity.  These modes
are characterized by low oscillation frequencies (long periods) and by
a  displacement of  the stellar  fluid essentially  in  the horizontal
direction  \cite{1989nos..book.....U, 1980tsp..book.....C}.   A single
best-fit   model   with    the   characteristics   shown   in   Tables
\ref{tab:G117-d} and  \ref{tab:R548-d} was  found for each  star.  The
second  column of  these  tables displays  the effective  temperature,
gravity  and stellar  masss ($T_{\rm  eff}$,  $\log g$  and $M_*$)  of
G117$-$B15A  and   R548,  respectively,  according   to  spectroscopic
studies.  The parameters characterizing the asteroseismological models
are shown  in column 3.   It is worth  noting the very  good agreement
between the spectroscopically inferred  values of $T_{\rm eff}$, $\log
g$ and $M_*$ and  the seismological results. These asteroseismological
models have proven be essential to  put constraints on the mass of the
axion   \cite{2012MNRAS.424.2792C,  2012JCAP...12..010C},  so   it  is
natural to use them to place constraints on a secularly varying $G$.

In  Tables~\ref{tab:G117-p}   and  \ref{tab:R548-p}  we   compare  the
observed and  theoretical periods and  rates of change of  the period.
Notably, the  models reproduce  very accurately the  observed periods.
This  guarantees  us  that   seismological  models  are  an  excellent
representation of the internal structure  of the real stars.  However,
the  most   important  fact   shown  in   Tables~\ref{tab:G117-p}  and
\ref{tab:R548-p} is  that the  observed rates of  change of  the 215~s
period of  G117$-$B15A and the  213~s period of  R548 are more  than 3
times larger  than the  theoretically expected  values.   This is
  because  the $\ell=  1, k=2$  mode is  a mode  trapped in  the outer
  hydrogen envelope in our  asteroseismological models.  Mode trapping
  reduces the rate of change of period  with time by up to a factor of
  $\sim 3$  if the  mode is  trapped in  the outer  hydrogen envelope,
  because gravitational  contraction --- that is  still appreciable in
  these  regions ---  reduces the  increase in  period due  to cooling
  \cite{B96}. Since the $\ell= 1, k  = 2$ mode is somewhat affected by
  gravitational  contraction,   it  is  less  sensitive   to  cooling.
  However, the change  of the period due to  the increasing degeneracy
  resulting  from cooling  is  still  larger than  the  change due  to
  residual contraction,  and so,  $\dot{\Pi} > 0$.   As a  result, the
  period of  the $\ell=1, k =  2$ mode is still  sensitive to cooling,
  and  therefore,  its  rate   of  change  reflects  the  evolutionary
  timescale  of the  stars.  Therefore,  the disagreement  between the
  observed and  theoretical values of  $\dot\Pi$ would be a  hint that
  G117$-$B15A and R548 could be  cooling faster than that predicted by
  the  standard  theory  of  white  dwarf  evolution.   This  possible
  anomalous  rate  of cooling  has  been  explained  in terms  of  an
  additional  cooling  produced   by  axions  \cite{Isern92,  Isern08,
    Isern10}.

\begin{table}
\centering
\caption{The  observed and theoretical  periods of  G117$-$B15A, along
  with the corresponding mode identification, and the subsequent rates
  of change  of period with  time. The theoretical ones  were computed
  assuming $\dot{G}= 0$.}
\begin{tabular}{cccccc}
\hline  
\hline
$\Pi^{\rm o}$        & $\Pi^{\rm t}$        & $\ell$ & $k$ & $\dot{\Pi}^{\rm o}$  & $\dot{\Pi}^{\rm t}$ \\  
$[$s$]$              &  $[$s$]$             &        &     & $[10^{-15} $s/s$]$   & $[10^{-15}  $s/s$]$ \\    
\hline  
  ---                &    189.19            &   1    &  1  & ---                  & 3.01                \\                
 215.20              &    215.22            &   1    &  2  & $4.19 \pm  0.53$     & 1.25                \\
 270.46              &    273.44            &   1    &  3  & ---                  & 4.43                \\
 304.05              &    301.85            &   1    &  4  & ---                  & 4.31                \\
\hline
\hline  
\end{tabular}  
\label{tab:G117-p}  
\end{table}  

\begin{table}  
\centering  
\caption{Same as Table \ref{tab:G117-p}, but for R548.}
\begin{tabular}{cccccc}  
\hline  
\hline  
$\Pi^{\rm o}$        & $\Pi^{\rm t}$       & $\ell$ & $k$ & $\dot{\Pi}^{\rm o}$  & $\dot{\Pi}^{\rm t}$ \\  
$[$s$]$              &  $[$s$]$            &        &     & $[10^{-15} $s/s$]$   & $[10^{-15}  $s/s$]$ \\    
\hline  
186.86               &   187.59            & 1      &  1  & ---                  & 2.51                \\  
212.95               &   213.40            & 1      &  2  & $3.3 \pm 1.1$        & 1.08                \\  
274.52               &   272.26            & 1      &  3  & ---                  & 3.76                \\  
318.08               &   311.36            & 2      &  8  & ---                  & 6.32                \\  
333.64               &   336.50            & 2      &  9  & ---                  & 8.80                \\  
\hline  
\hline  
\end{tabular}  
\label{tab:R548-p}  
\end{table}  

%______________________________________________________________________ 

\section{The rates of period change with a varying $G$}
\label{method}

In the  previous section we  have presented  the periods and  rates of
period change of G117$-$B15A and R548  that do not take into account a
possible secular rate  of variation of the  gravitational constant $G$
--- see Tables  \ref{tab:G117-p} and \ref{tab:R548-p}.  We  have found
that the  theoretically expected  rate of period  change of  the modes
with  $k  =  2$  is  markedly smaller  than  the  value  measured  for
G117$-$B15A  and R548,  suggesting  the existence  of some  additional
cooling mechanism in these stars. In this section we shall assume
  that this additional  cooling is entirely due to a  varying $G$.  In
  this regard  we point  out that  the upper  bounds derived  here are
  conservative, since we attribute  the entire discrepancy between the
  observed and the theoretical cooling rates of these white dwarfs ---
  see Sect.~\ref{UN} --- to a varying  $G$, which is an extreme case.
We will moreover assume that $\dot{G} < 0$, unless otherwise stated.

\begin{figure}[t]
\centering
\includegraphics[width=0.8\textwidth]{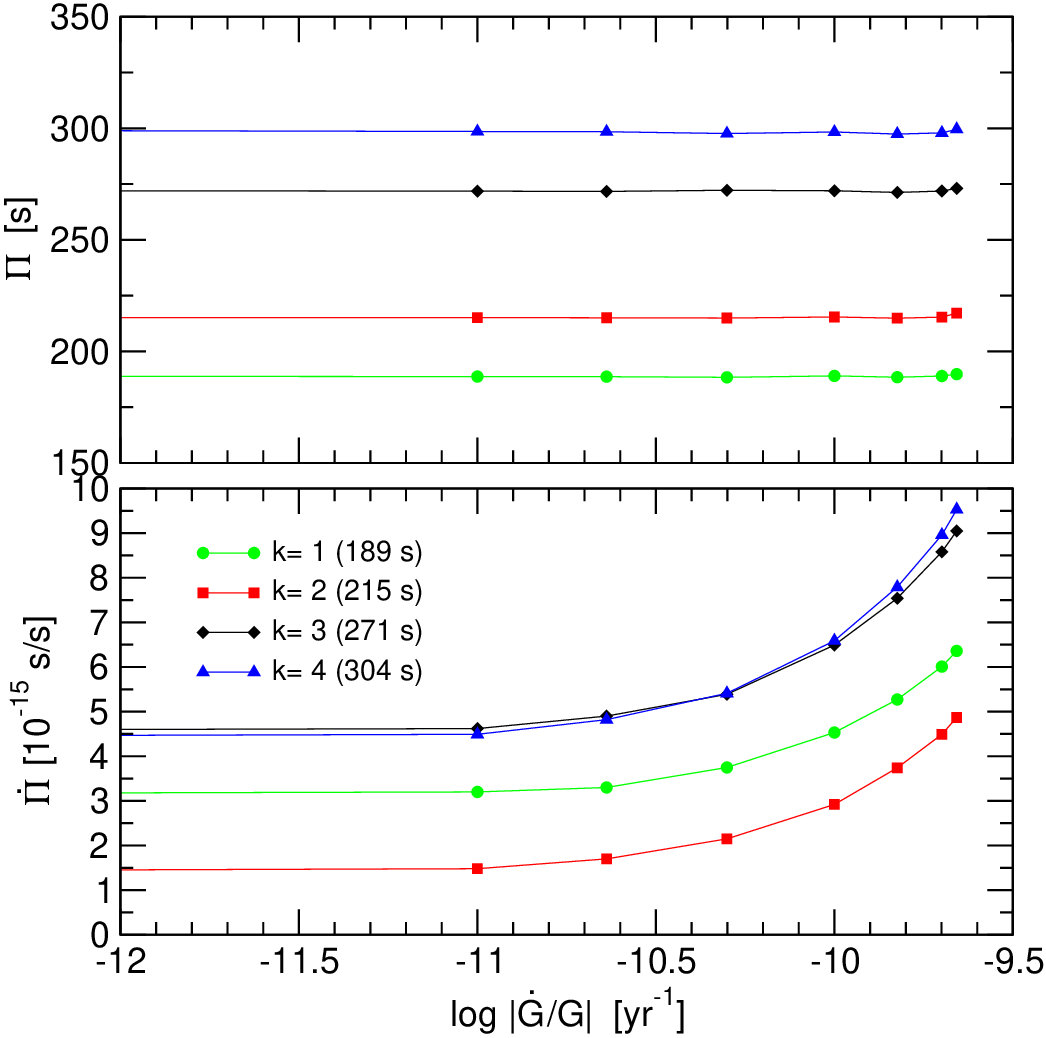}
\caption{Upper panel: periods of the several modes of G117$-$B15A as a
  function  of the value  of $|\dot  G/G|$ with  $\dot{G} <  0$. Lower
  panel: period  derivatives of  the same modes  as a function  of the
  secular  rate of  change of  $G$. The  effective temperature  of the
  models is $T_{\rm eff} \sim 11\,990$ K.  See text for details.}
\label{fig:periods}
\end{figure}

Following Ref.~\cite{gnew}, we  have computed a set of  DA white dwarf
cooling  sequences  considering  a  secularly  varying  $G$.   In  the
interest  of  simplicity  we  have assumed  that  $\dot{G}/G$  remains
constant with  time. We have adopted different  values for $\dot{G}/G$
and  the  same structural  parameters  ($M_*,  M_{\rm H}$)  previously
obtained  in our  asteroseismological  models.  We  have considered  a
range of values  for the rate of change  of the gravitational constant
$5\times  10^{-12}\ {\rm  yr}^{-1}  \leq |\dot{G}/G|  \leq 2.5  \times
10^{-10}\ {\rm yr}^{-1}$.  It  is worth emphasizing that the evolution
with a varying  $G$ is strongly dependent on the  initial value of $G$
at the  beginning of the cooling phase  \cite{gnew}.  Accordingly, for
each value of $\dot{G}/G$ we computed several sequences with different
values of $G_{\rm  i}/G_0$, where $G_{\rm i}$ stands  for the value of
$G$ at time $t_{\rm i}$, corresponding to the beginning of the cooling
track  at high  effective temperature,  and $G_0$  corresponds  to the
present  value  of  $G$.   Specifically,  we  have  assumed  that  the
variation of $G$ is exponential,  and follows a simple exponential law
of the form:

\begin{equation}
G(t)= G_i \exp \left[\frac{\dot{G}}{G}(t-t_i)\right],
\end{equation}

To  obtain starting  white dwarf  configurations with  the appropriate
value of $G$, we simply changed  $G_0$ by $G_{\rm i}$ at the beginning
of  the  cooling  phase  (at  time $t_{\rm  i}$)  and  calculated  the
resulting  structure. After a  brief transitory  stage, we  obtain the
initial configurations for our sequences. This artificial procedure to
obtain the  initial white dwarf configurations is  justified since the
subsequent  evolution does  not depend  on the  way the  initial white
dwarf structures are obtained. Note that the election of the values of
$G_{\rm  i}$ was  made in  such  a way  that $G(t)\equiv  G_0$ at  the
effective   temperature   of    these   stars   according   to   their
asteroseismological models,  that is, $T_{\rm eff} \sim  11\,990$ K in
the case of G117$-$15A, and $T_{\rm  eff} \sim 11\,630$ K for R548. In
this way, we consistently assure that  the varying value of $G$ in our
evolutionary computations  is coincident at the present  time with the
present value of the gravitational constant.
 
The pulsation periods for the modes with  $\ell = 1$ and $k = 1, 2, 3$
and $4$ of the asteroseismological model of G117$-$B15A for increasing
values  of   $|\dot{G}/G|$  are  depicted   in  the  upper   panel  of
Fig.~\ref{fig:periods}.  The  variation of the  periods is negligible,
in  spite  of  the  rather  wide  range  of  values  of  $|\dot{G}/G|$
considered  here ---  spanning  about two  orders  of magnitude.  This
result  implies that  due to  a  varying $G$  with $\dot{G}  < 0$  the
structure  of  the  asteroseismological  model  itself  is  negligibly
affected,  in such  a way  that, for  a fixed  value of  the effective
temperature,  the pulsation  periods  are largely  independent of  the
adopted value of $|\dot{G}/G|$.

\begin{figure}[t]
\centering
\includegraphics[width=0.8\textwidth]{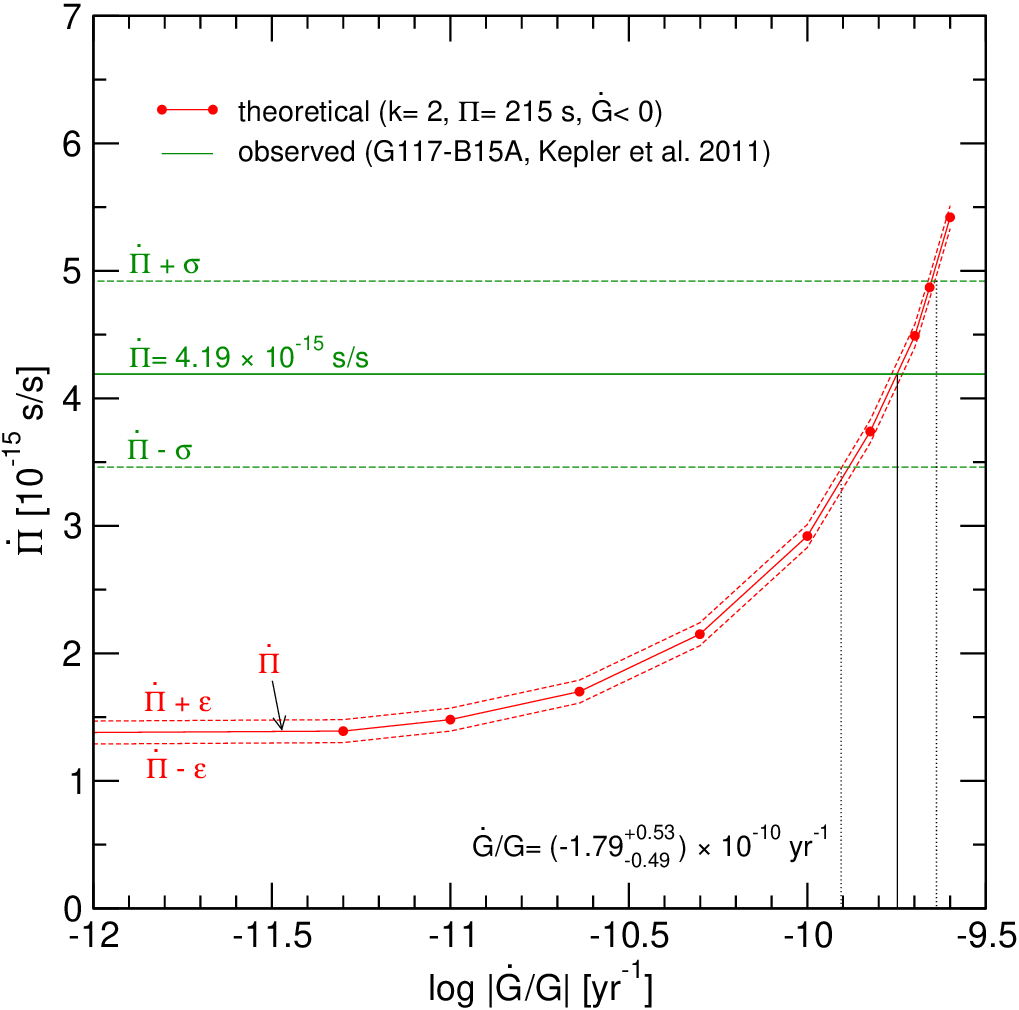}
\caption{Rate of temporal variation of the 215~s period of G117$-$B15A
  as  a function of  the value  of $\dot{G}/G$,  red solid  line.  The
  effective temperature  of the models  is $T_{\rm eff}  \sim 11\,990$
  K. The observational  value of the rate of change  of the period ---
  horizontal solid  line ---  along with its  observed error  bars ---
  horizontal  dashed   lines  ---  is  also   displayed.   The  formal
  theoretical errors are also shown as dashed lines.}
\label{fig:G117}
\end{figure}

In the lower  panel of Fig.~\ref{fig:periods} we display  the rates of
period change for  the same modes. At odds with  what happens with the
pulsation  periods, the  values  of  the rates  of  period change  are
markedly  affected  by a  varying  $G$,  substantially increasing  for
increasing  values of  $|\dot{G}/G|$.   This is  because  that, for  a
decreasing value  of $G$  with time, the  white dwarf  cooling process
accelerates \cite{gold,  gnew}, and this  is translated into  a larger
secular change of the pulsation periods as compared with the situation
in which $G$ is constant. In  particular, the rate of period change of
the  mode with  $k =  2$, which  is the  relevant one  in  the present
analysis,  increases  by  a  factor   of  about  5  in  the  range  of
$|\dot{G}/G|$ considered in the case $\dot{G}<0$.

For  sake  of conciseness,  we  have  shown  results corresponding  to
G117$-$B15A  only, but  the same  analysis can  be done  for  R548 and
identical results are obtained.  Finally, we mention that we have also
computed  a  set  of  models  in  which  we  vary  $G$  but  we  adopt
$\dot{G}>0$. In this case, we  found that the pulsation periods do not
change, and  that the rates  of period change decrease  for increasing
values of  $|\dot{G}/G|$. This case is  not of direct  interest in the
context of this paper, because to reconcile the the difference between
the theoretical and the measured  values of the rates of period change
in the ZZ Ceti stars G117$-$B15A and R548 a physical mechanism able to
increase these rates  must be sought.  Thus, we  will not consider the
case of a varying $G$ with $\dot{G}>0$ in the remainder of the paper.

%______________________________________________________________________ 

\section{An upper limit to the rate of variation of $G$}
\label{G}

\begin{figure}[t]
\centering
\includegraphics[width=0.8\textwidth]{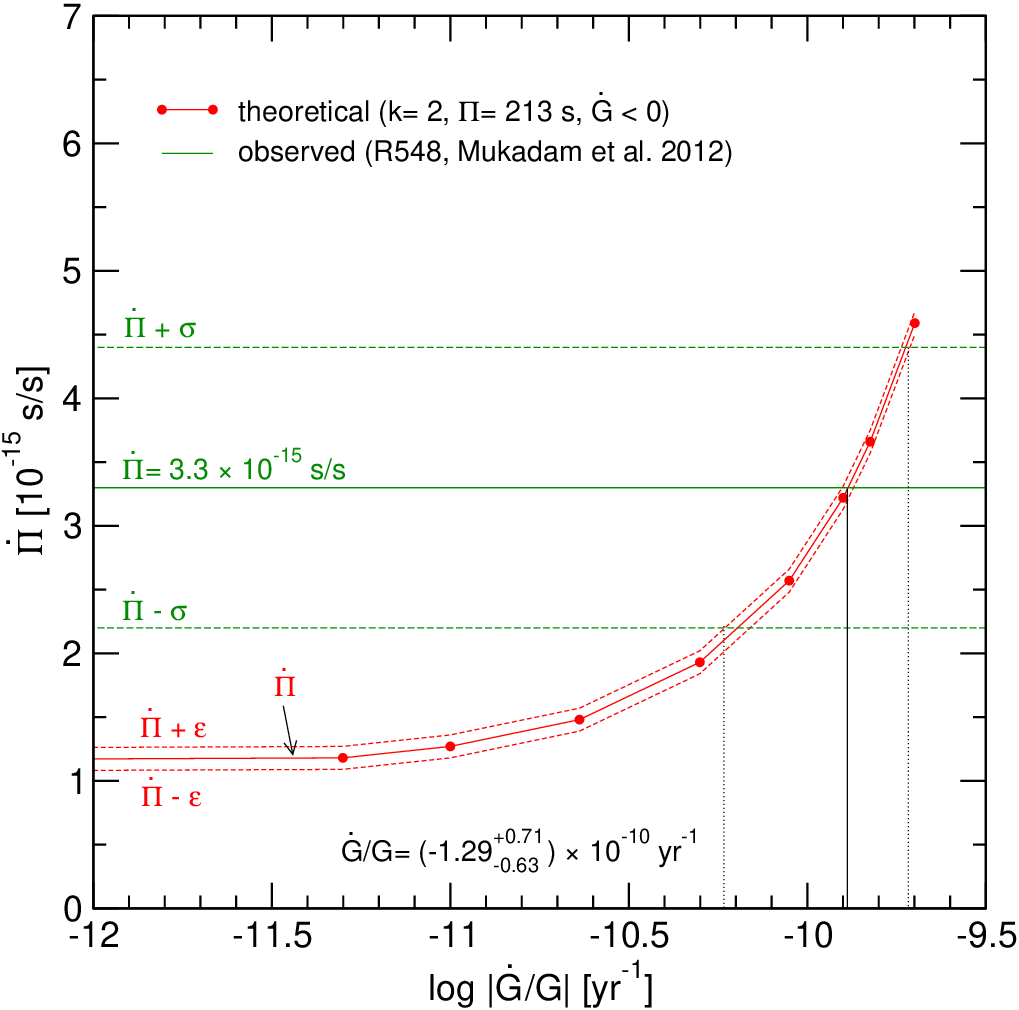}
\caption{Same   as  Fig.~\ref{fig:G117}   for   R548.  The   effective
  temperature of the models is $T_{\rm eff} \sim 11\,630$ K.}
\label{fig:R548}
\end{figure}

Here,  we focus  on  the  mode with  $k  = 2$,  for  which  we have  a
measurement  of its  rate of  period change  for both  G117$-$B15A and
R548.   In Fig.  \ref{fig:G117}  we display  the theoretical  value of
$\dot{\Pi}$ corresponding  to the period $\Pi =  215$~s of G117$-$B15A
for increasing  values of $|\dot{G}/G|$ (red solid  curve). The dashed
curves  embracing the  solid curve  represent the  uncertainty  in the
theoretical value of  $\dot{\Pi}$, $\epsilon_{\dot{\Pi}} = 0.09 \times
10^{-15}$  s s$^{-1}$. This  value has  been obtained  considering the
uncertainty  introduced  by  our  lack  of precise  knowledge  of  the
$^{12}$C$(\alpha,\gamma)^{16}$O reaction  rate --- $\varepsilon_1 \sim
0.03  \times  10^{-15}$~s/s,  and  that  due  to  the  errors  in  the
asteroseismological   model  ---   $\varepsilon_2  \sim   0.06  \times
10^{-15}$~s/s  \cite{2012MNRAS.424.2792C}. We  are  assuming that  the
uncertainty for the case in which $\dot{G} \neq 0$ is the same as that
computed for  the case in which  $G= 0$.  If we  consider one standard
deviation from  the observational value, we conclude  that the secular
rate of  variation of the gravitational constant  using G117$-$B15A is
$\dot{G}/G=  (-1.79^{+0.53}_{-0.49})  \times 10^{-10}$~yr$^{-1}$.  The
same analysis applied  to R548 --- see Fig.   \ref{fig:R548} --- gives
$\dot{G}/G= (-1.29^{+0.71}_{-0.63}) \times 10^{-10}$~yr$^{-1}$.  These
values ​​are completely compatible each other, although currently less
restrictive than those obtained using other techniques.

\section{Discussion}
\label{UN}

We emphasize  at this point that the enhanced  rate of cooling of
  G117$-$B15A and R548 --- which is assumed here to be entirely due to
  a varying $G$ --- and the determination of a new upper bound for the
  variation of $G$ is based on the  fact that our set of full DA white
  dwarf evolutionary  models predicts that the  $\ell= 1, k =  2$ mode
  ($\sim 215$~s period  for G117$-$B15A and $\sim 213$~s  for R548) is
  strongly  trapped  in  the  outer   H  envelope  of  the  respective
  asteroseismological models.  Had this  mode not been largely trapped
  in the  outer layers,  larger values  of its  rate of  period change
  would have  resulted \cite{2012MNRAS.424.2792C}.  In this  case, the
  upper bound on  $\dot{G}/G$ should have been tighter.   These can be
  estimated by considering the value of the rate of period change that
  the $\ell= 1, k = 2$ mode would have if it were non-trapped.  A very
  simple way of  doing this is to  assume that the mode  should have a
  $\dot{\Pi}$  similar to  the typical  value  of the  rate of  period
  change  of the  dipole  low-order non-trapped  modes,  that is  with
  $\ell= 1$ and $k = 1, 3$ and  $4$ for G117$-$B15A, and $\ell= 1, k =
  1$ and $3$ for R548.  We obtain $\dot{\Pi}^{\rm t}_{\rm nt} \sim 3.9
  \times  10^{-15}$~s/s,  and  $\dot{\Pi}^{\rm t}_{\rm  nt}  \sim  3.1
  \times   10^{-15}$~s/s,   respectively.     Thus,   the   value   of
  $\dot{\Pi}^{\rm  t}$  for  the  $\ell=  1,   k  =  2$  mode  of  the
  asteroseismological model  could be  $\sim 2.7  \times 10^{-15}$~s/s
  larger for  G117$-$B15A, and  $\sim 2.1 \times  10^{-15}$~s/s larger
  for R548, if it were a  non-trapped mode.  Taking this into account,
  we obtain  $|\dot{G}/G| \lesssim  10^{-11}$~yr$^{-1}$, in  line with
  other independent  upper limits, but  our result is  consistent with
  $\dot{G}= 0$.

  Another point  that is  worth discussing  is the  fact that  we have
  considered  only  internal  errors   associated  to  the  period-fit
  procedures  that   lead  to   the  asteroseismological   models  for
  G117$-$B15A  and R548.   However, it  is important  to realize  that
  substantially  different  asteroseismological fits  (with  different
  values for $M_*$, $T_{\rm eff}$,  $M_{\rm H}$,\ldots) are found when
  different  sets  of  DA  white  dwarf models  are  employed  in  the
  analysis,  being the  corresponding  values of  $\dot{\Pi}$ for  the
  $\ell= 1,  k= 2$ mode quite  different from those obtained  here for
  G117$-$B15A and  R548.  Specifically, the spread  in the theoretical
  value  of $\dot{\Pi}$  for  the 215~s  period  according to  already
  published asteroseismological  studies on G117$-$B15A  is relatively
  large:       $\dot{\Pi}=        3.7       \times       10^{-15}$~s/s
  \cite{1998ApJS..116..307B},  $\dot{\Pi}=  3.9  \times  10^{-15}$~s/s
  \cite{2001NewA....6..197C},    and     $\dot{\Pi}=    2.98    \times
  10^{-15}$~s/s       and       $1.92       \times       10^{-15}$~s/s
  \cite{2008ApJ...675.1512B}.   Note,  nevertheless,  that  all  these
  values are  smaller than the  observed one ($\dot{\Pi}=  4.19 \times
  10^{-15}$~s/s).   Hence, all  the  existing  analysis indicate  that
  G117$-$B15A cools faster  than expected, although the  spread in the
  value of  $\dot{\Pi}$ for the 215~s  period is rather large  $\sim 2
  \times  10^{-15}$~s/s. This  spread can  be adopted  as a  realistic
  value of the uncertainty in the theoretical rate of period change of
  this  mode.   Taking into  account  this  uncertainty, that  clearly
  dominates over all the errors  affecting the $\dot{\Pi}^{\rm t}$, we
  found that  $10^{-12}   \lesssim   |\dot{G}/G|   \lesssim   3   \times
  10^{-10}$~yr$^{-1}$, being  the lower  limit in very  good agreement
  with  the  results  obtained   using  other  techniques.   The  same
  conclusion can be reached when R548 is considered.

%______________________________________________________________________ 

\section{Summary and conclusions}
\label{conc}

In this paper,  we have used an  state-of-the-art stellar evolutionary
code and  an up-to-date pulsational code  to compute the effects  of a
secularly  varying  gravitational  constant  $G$  on  the  pulsational
properties of  variable white dwarfs.  Specifically,  we have compared
the theoretical  rates of period  change obtained taking  into account
the effects of a running $G$ with  the measured rates of change of the
periods of  two well studied  pulsating white dwarfs,  G117$-$B15A and
R548.  We  assumed that  $\dot{G}/G$ remains  constant with  time, and
considered a rate of change of the gravitational constant in the range
$5\times  10^{-12}\ {\rm  yr}^{-1}  \leq |\dot{G}/G|  \leq 2.5  \times
10^{-10}\ {\rm yr}^{-1}$.  We found  that the pulsation periods do not
experience appreciable  changes, but  the rates  of period  change are
strongly affected  when a  varying $G$ is  assumed. In  particular, we
found  that  the rates  of  period  change increase  when  $\dot{G}/G$
increases if $\dot{G}<0$, and the opposite  holds in the case in which
$\dot{G}>0$. As a result of our analysis, we have found an upper bound
$|\dot G/G|\sim 1.8\times 10^{-10}$~yr$^{-1}$ using the variable white
dwarf G117$-$B15A,  and $|\dot G/G|\sim  1.3\times 10^{-10}$~yr$^{-1}$
using R548.  These  bounds are considerably weaker  than those derived
from   Lunar  Laser   Ranging  ---   $\dot{G}/G  =   (0.2\pm0.7)\times
10^{-12}$~yr$^{-1}$ \cite{H10}  --- and  Big Bang  nucleosynthesis ---
$-0.3   \times   10^{-12}$~yr$^{-1}   \la  \dot{G}/G   \la   0.4\times
10^{-12}$~yr$^{-1}$ \cite{CO4,B05}  --- but as stringent  as the upper
limits inferred  from Hubble diagram  of Type Ia supernovae  --- $\dot
G/G\la 1\times 10^{-11}$~yr$^{-1}$~\cite{SNIa,IJMPD}.  Our upper bound
on the secular rate of variation  of $G$ is also less restrictive than
that obtained  from solar  helioseismology --- $\dot  G/G\la 1.6\times
10^{-11}$~yr$^{-1}$~\cite{Guenther98}.   In any case, we emphasize
  that  while the  limits reported  here are  not as  strict as  those
  obtained  using other  methods, they  are independent  and therefore
  rest on a different set of model dependent parameters.

 As mentioned, our results rely on  the fact that the $\ell= 1, k=
  2$   modes  are   trapped   in   the  outer   H   envelope  of   the
  asteroseismological models for G117$-$B15A and  R548.  As it is well
  known, mode  trapping in white  dwarfs is strongly sensitive  to the
  precise  shape  of  the   chemical  profiles.   Since  the  chemical
  structure of  white dwarfs  is constructed  during the  evolution of
  their  progenitor stars,  a  full exploration  of the  uncertainties
  affecting the white dwarf and pre-white dwarf modelling is needed to
  investigate if the  $\ell= 1, k= 2$ mode  in the asteroseismological
  models is indeed a trapped one.

In closing, we  mention that the bounds obtained  here using pulsating
white  dwarfs could  be improved  if the  rate of  period change  were
measured  for a  massive pulsating  white dwarf.  This is  because the
effects  of a  varying $G$  are stronger  for the  more massive  white
dwarfs \cite{gnew}.  Actually, it is  expected that the enhancement of
the cooling in a massive pulsating white dwarf should translate into a
larger increase  of its rates  of period  change, thus resulting  in a
more stringent upper limit on $\dot{G}/G$.

\acknowledgments This  research was  supported by AGENCIA  through the
Programa  de  Modernizaci\'on  Tecnol\'ogica  BID 1728/OC-AR,  by  PIP
112-200801-00940 grant from CONICET, by MCINN grant AYA2011--23102, by
the ESF EUROCORES Program EuroGENESIS (MICINN grant EUI2009-04170), by
the European Union FEDER funds, and by the AGAUR.

\bibliographystyle{JHEP}
\bibliography{varG-r1}

\end{document}